\documentstyle[11pt,newpasp,twoside,epsf]{article}
\def\edcomment#1{\iffalse\marginpar{\raggedright\sl#1\/}\else\relax\fi}
\reversemarginpar

\begin{document}
\title{HST Observations of Young Stellar Clusters in Nearby Galaxies}
 \author{S{\o}ren S. Larsen}
\affil{European Southern Observatory, Karl-Schwarzschild-Str.\ 2, D-85748
       Garching bei M{\"u}nchen, Germany}

\begin{abstract}
  The HST data archive contains images of several 
nearby spirals, suitable for detailed studies of the properties of 
individual star clusters and their surroundings. By combining 
structural information derived from HST images with ground-based 
photometry, it is possible to study cluster properties as a
function of age and mass. While both the core- and effective radii of
young clusters correlate with mass, the slopes of these relations 
are shallower than for a constant-density relation, implying that the 
mean density increases with cluster mass. This must be accounted for
in theories for cluster formation.
\end{abstract}

\section{Introduction}

  While the rich cluster systems in merger galaxies and starbursts
have been studied intensively over the past decade, it is sometimes
overlooked that even some ``normal'', undisturbed spiral galaxies
can form highly luminous, young star clusters (Larsen \& Richtler 1999;
Paper I).  Within a distance of 10 Mpc, there are several examples of 
spirals with rich cluster systems which have been imaged with HST/WFPC2,
allowing the properties
and environment of individual star clusters to be studied in great detail.
By combining the HST images with ground-based photometry, it is
possible to look for trends in structural parameters with cluster age 
and mass, providing potentially valuable clues to the physics of 
cluster formation and subsequent dynamical evolution.

\section{Cluster structure versus mass and age}

  We have recently scanned the HST archive for images of galaxies
included in the original survey from Paper I. A total of 17 galaxies 
were found to have adequate HST data (Larsen 2004). The constraint that the 
clusters have to be visible on ground-based images is not as severe as 
it might appear at first, since a high S/N in the HST images is anyway 
required to measure accurate structural parameters. By 
fitting ``Moffat'' models of the form $P(r) \propto [1+(r/r_c)^2]^{-\alpha}$, 
it is possible to constrain both the core radius $r_c$ and the shape 
of the luminosity profile at large radii, parameterised by the 
slope $\alpha$. While these profiles have no physical motivation, they have
been shown to fit LMC clusters well (Elson et al.\ 1987).
For $\alpha=1$, they are identical to a King (1962) model with
infinite concentration parameter.

  The left panel in Fig.~1 shows the distribution of envelope slopes
for clusters in four age bins. There is a clear tendency for many of the
youngest clusters to have extended outer envelopes. This may reflect the 
density structure of the parent molecular cloud cores, or be a hint that
some of these clusters are unbound and expanding.  Older clusters 
gradually evolve towards a King-like profile.  In the right-hand panel, the 
FWHM and half-light radii are shown versus cluster mass.  In each 
panel, the constant-density relation (size $\propto M^{1/3}$) is shown with 
a dashed line, while the solid lines indicate a fit to the data. For the 
half-light radii, only clusters with $1<\alpha<5$ were included in the fit 
(clusters with $\alpha<1$ have poorly defined half-light radii and are shown 
with diamond symbols). Although both relations show substantial scatter, 
the fits are significantly shallower than for the constant-density 
relation. This conclusion is robust to selection effects, which would 
be expected to produce a bias against extended, low-mass objects.

  The main limitation in this study is the difficulty of detecting
low-mass clusters older than a few times $10^7$ years. This situation
is expected to improve when new ACS data become available.

\begin{figure}
\plottwo{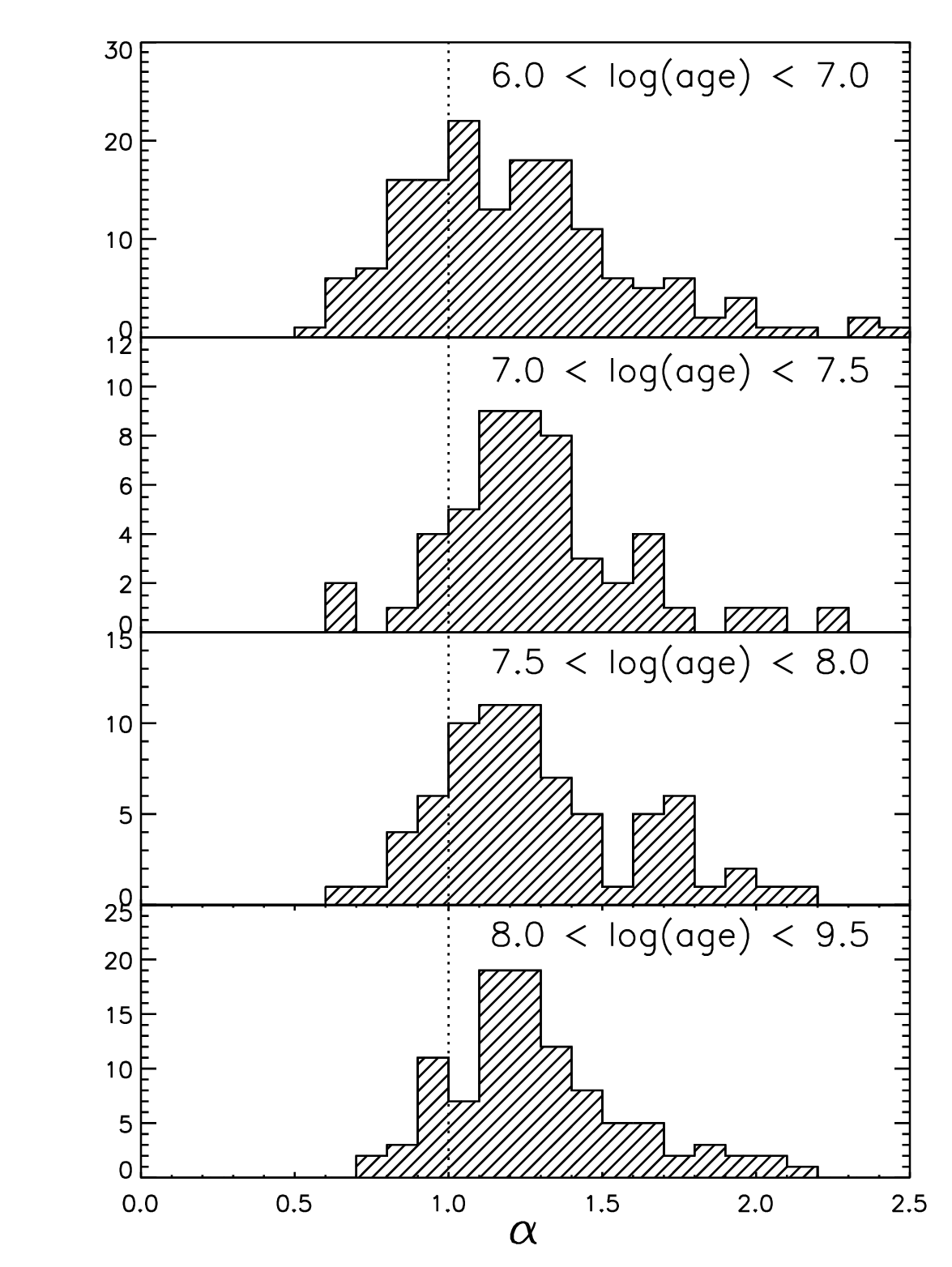}{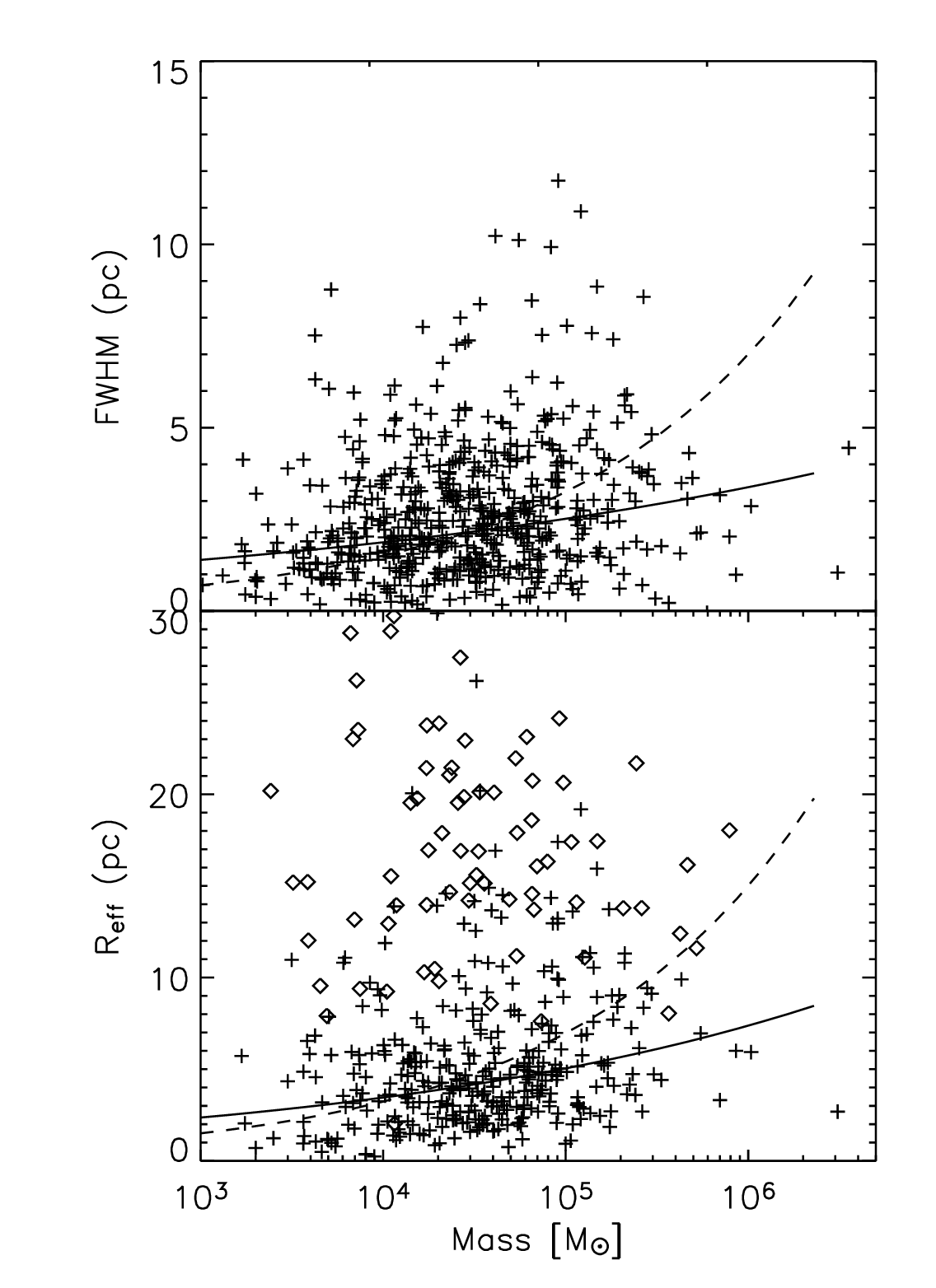}
\caption{Left: Distribution of envelope slopes for clusters in 
  four age bins. Right: FWHM and half-light radius versus mass.}
\end{figure}


\begin{references}
  \reference Elson, R.\ A.\ W., Fall, S.\ M.\ \& Freeman, K.\ C.\ 1987, 
             \apj, 323, 54
  \reference King, I.\ R.\ 1962, \aj, 67, 471
  \reference Larsen, S.\ S.\ \& Richtler, T.\ 1999, \aap, 345, 59
  \reference Larsen, S.\ S.\ 2004, \aap, in preparation
\end{references}
\end{document}